\documentclass[conference]{IEEEtran}
\IEEEoverridecommandlockouts

\usepackage{cite}
\usepackage{amsmath,amssymb,amsfonts}
\usepackage{algorithmic}
\usepackage{graphicx}
\usepackage{textcomp}
\usepackage{xcolor}
\usepackage{multirow}

\def\BibTeX{{\rm B\kern-.05em{\sc i\kern-.025em b}\kern-.08em
    T\kern-.1667em\lower.7ex\hbox{E}\kern-.125emX}}
\begin{document}

\title{Cross-Dataset, Age, and Gender Generalization: A Comprehensive Analysis of Fine-Tuning Strategies for Low-Resource Children’s ASR}

\author{\IEEEauthorblockN{Abhijit Sinha$^{1}$, Hemant Kumar Kathania$^{1}$, Sudarsana Reddy Kadiri$^{2}$, and Shrikanth Narayanan$^{2}$}
\IEEEauthorblockA{\textit{$^{1}$Department of Electronics and Communication Engineering, National Institute of Technology Sikkim, India} \\
\textit{$^{2}$Signal Analysis and Interpretation Laboratory, University of Southern California, Los Angeles, USA}\\
}}


\maketitle

\begin{abstract}

Children’s speech recognition remains challenging due to acoustic variability, dataset mismatches, and pretraining biases. Self-supervised models like Wav2Vec2 and HuBERT have achieved strong performance in adult ASR, but adapting these models to low-resource children’s speech remains limited. In this study, we evaluate age-specific, gender-specific, and cross-dataset fine-tuning strategies on the PFSTAR and CMU Kids datasets. Our findings reveal three key patterns: (i) \textit{fine-tuning on younger children’s speech improves generalization to older children’s speech}, (ii) \textit{fine-tuning reduces male-preference bias}, and (iii) \textit{cross-dataset performance drops significantly due to accent and vocabulary mismatches}. Notably, the shorter utterances in the CMU Kids corpus lead to higher baseline WER, highlighting that SSL models struggle with brief speech without adaptation. These insights provide actionable guidelines for developing robust and inclusive ASR systems for children, emphasizing that child-centric ASR benefits from targeted fine-tuning and diverse pretraining data.

\end{abstract}


\section{Introduction}
\label{sec:intro}
Automatic Speech Recognition (ASR) systems have revolutionized human-computer interaction, enabling applications such as voice assistants, educational tools, and accessibility services. Despite these advances, accurate ASR for children remains a formidable challenge due to the distinct acoustic and linguistic properties of young speakers. Compared to adult speech, children’s speech exhibits a higher pitch, greater variability in pronunciation, and faster speaking rates \cite{lee1999acoustics,Vorperian2007VowelAS}. These factors contribute to significantly elevated error rates when ASR systems typically trained on extensive adult corpora are applied to children \cite{koenig2008speech,10.21437/interspeech.2020-3037,yeung2018difficulties}. Compounding the problem, there is a scarcity of large, labeled datasets for children’s speech \cite{claus2013survey,feng2024towards,Sukhadia2024ChildrensSR}, which limits the models’ ability to capture the full range of variability across different ages and speech patterns.

To mitigate these challenges, researchers have explored a range of techniques. Data augmentation methods such as time scale modification \cite{Fan2023UsingMA,Shahnawazuddin2022ImprovingTP,sinha2024effect}, formant modification \cite{kathania_2020}, and vocal tract length normalization \cite{Patel2024ImprovingEM} have been proposed to artificially increase the diversity of training data. Additionally, transfer learning \cite{Rolland2022MultilingualTL,Thienpondt2022TransferLF} and domain adaptation \cite{Fan2022DRAFTAN} strategies have been utilized to fine-tune models-originally trained on large adult datasets-with limited child-specific data \cite{russell2006pf,eskenazi1997cmu}. More recently, self-supervised learning (SSL) has emerged as a powerful approach for ASR. SSL models such as Wav2Vec2 \cite{baevski2020wav2vec}, HuBERT \cite{hsu2021hubert}, Data2Vec \cite{baevski2022data2vec}, and WavLM \cite{chen2022wavlm} can learn robust speech representations from vast amounts of unlabeled audio, and studies have shown that fine-tuning these models on children’s speech significantly improves performance \cite{Fan2022DRAFTAN,fan2022towards,jain2023wav2vec2,Li2024AnalysisOS}.

However, most prior work has focused on overall accuracy and has not fully addressed the inherent variability within the child population. ASR generalization across age groups, genders, and datasets remains a key limitation. Younger children’s speech often exhibits higher acoustic variability, which can impact cross-age performance. Similarly, gender-related differences-such as pitch and articulation-can introduce biases from adult pretraining that affect accuracy. Cross-dataset generalization is also a challenge: recognition WER often increases substantially on corpora with different accents or demographics. Although previous studies showed that fine-tuning on children’s speech can reduce some pretraining biases, the specific impacts of age, gender, utterance length, and vocabulary complexity remain under-explored.

Motivated by these challenges, our study is the first to systematically analyze age- and gender-specific fine-tuning of SSL models for children’s ASR in low-resource scenarios. Specifically, we investigate:

\begin{itemize}

    \item How acoustic diversity in younger children’s speech influences generalization to older children.
    \item Pretraining biases revealed by gender-specific fine-tuning, which may favor male speech.
    \item Challenges of cross-dataset generalization, where performance degrades significantly due to accent and demographic mismatches.
\end{itemize}

We conduct experiments on two well-established children’s speech corpora, PFSTAR \cite{russell2006pf} and CMU Kids \cite{eskenazi1997cmu}, using three pre-trained SSL models (Wav2Vec2, HuBERT, and WavLM) under age-specific, gender-specific, and cross-dataset fine-tuning. Our results indicate that models fine-tuned on younger children generalize better to older children’s speech, fine-tuning reduces male speech bias, and shorter utterances suffer higher WER, highlighting limitations of current SSL architectures on brief speech. These findings provide practical guidelines for developing robust, child-centric ASR systems. Overall, our study offers insights to guide the design of ASR systems that are both robust and inclusive of children’s diverse speech characteristics.

\section{Experimental Framework}

\begin{figure}[!h]
\centering
\includegraphics[width=0.49\textwidth]{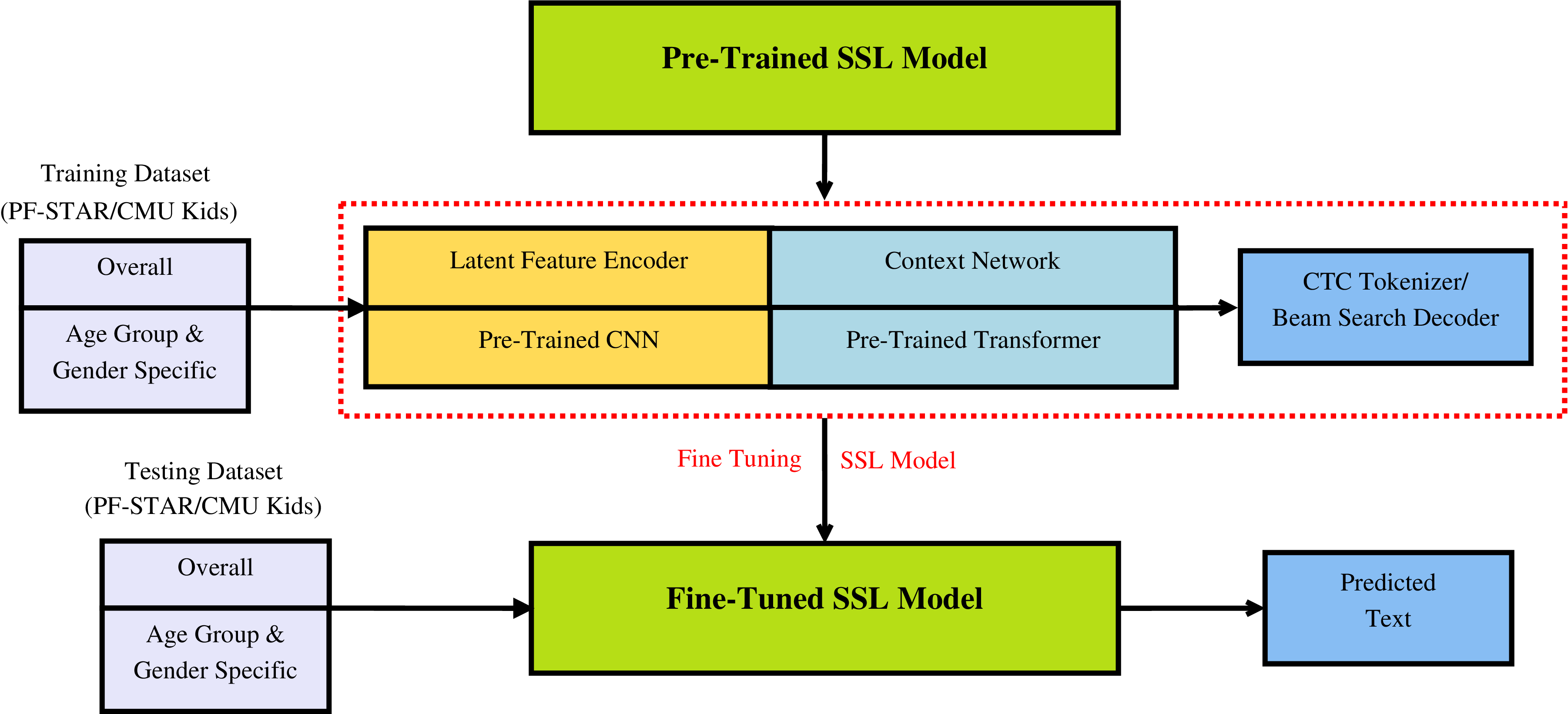}
\caption{A schematic block diagram illustrating the fine-tuning of SSL models on children's speech data, including both the overall training set and age-group specific subsets, followed by testing on the corresponding test set and its age-group specific subsets.}
\label{fig:wav2vec2}
\end{figure}

Figure \ref{fig:wav2vec2} provides a schematic overview of the fine-tuning process for SSL models in the context of children’s ASR. The process begins with a pre-trained SSL model such as Wav2Vec2, HuBERT, or WavLM-trained on large, general speech datasets. This model is then fine-tuned on children’s speech data, tailored to specific age-group and gender subsets. Fine-tuning adjusts the model’s parameters so that it better captures the unique characteristics of children’s speech, which differ from adult speech in terms of pitch, speaking rate, and pronunciation variability.

\vspace{-5pt}

The architecture of these SSL models comprises two main stages. First, a convolutional neural network (CNN) extracts features from raw speech signals, converting them into a sequence of feature vectors. Second, a Transformer-based context network processes these vectors to capture long-range dependencies and temporal relationships in the speech signal. The attention mechanisms within the Transformer allow the model to focus on the most relevant parts of the input sequence, thereby learning contextual patterns and nuances in children’s speech.

A critical component of SSL is the masking mechanism employed during training. Portions of the input speech features are randomly masked, and the model is tasked with predicting the missing information based on the surrounding context. This self-supervised learning strategy enables the construction of robust speech representations without relying on extensive labeled data, which is particularly beneficial in low-resource scenarios involving children’s speech.

\section{Datasets and Experimental Setup}
\label{sec:data} 

\subsection{Datasets}

This study utilizes two well-known children’s speech datasets: PFSTAR \cite{russell2006pf} and CMU Kids \cite{eskenazi1997cmu}. The PFSTAR dataset consists of British English recordings of children aged 4 to 14 years, with 8.3 hours of training speech from 122 speakers and 1.1 hours of testing speech from 60 speakers. In contrast, the CMU Kids dataset contains American English recordings of children reading sentences, with ages ranging from 6 to 11 years. It features 5180 utterances from 76 speakers, of which 70\% (6.3 hours) is used for training and the remaining 30\% (2.83 hours) for testing.

\begin{figure}[!h]
    \centering
    \includegraphics[width=8.6cm,height=4.6cm]{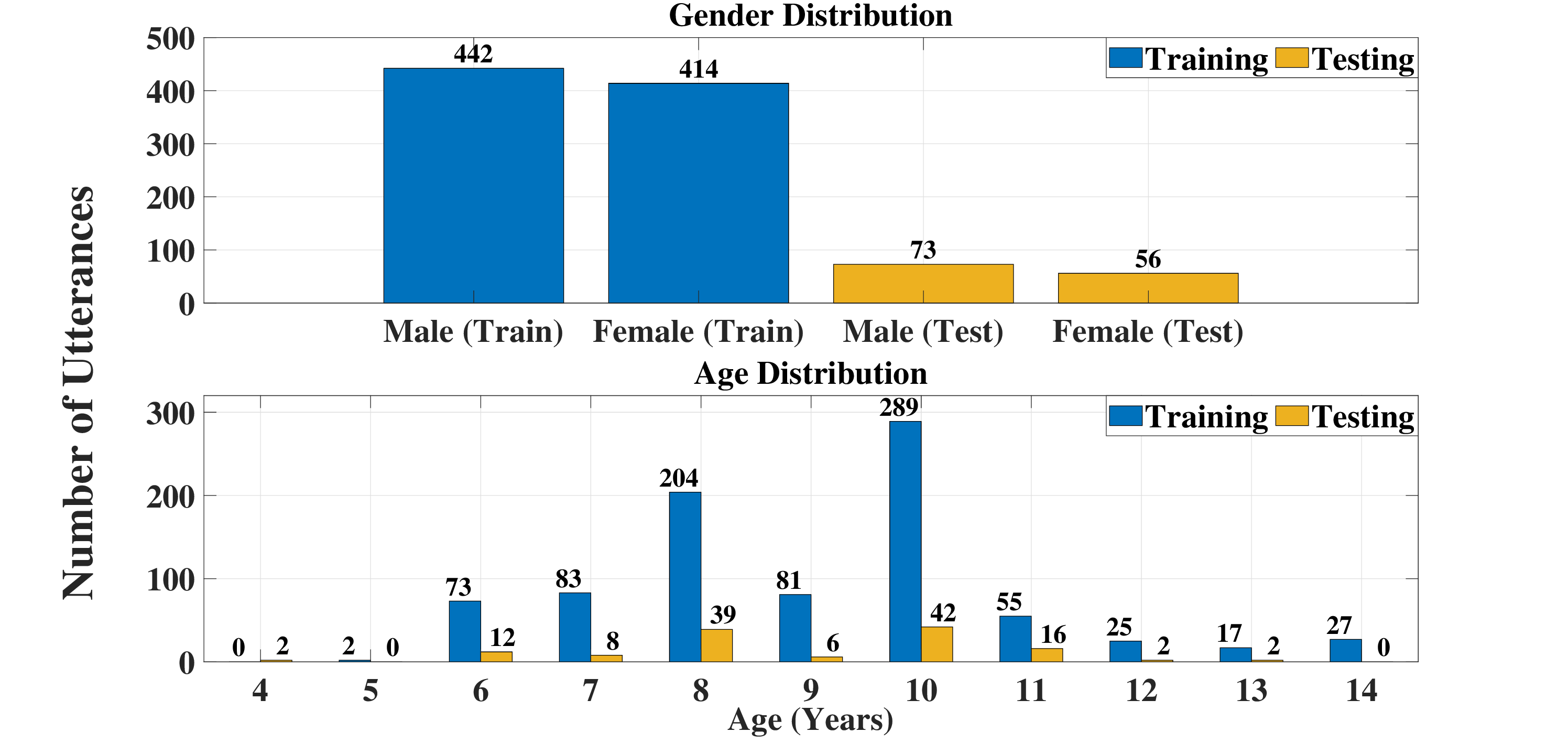} 
    
    \caption{Age and gender distribution of the PFSTAR dataset, showing training and testing splits by speaker gender and age.}
    \label{fig:data_dist_PFSTAR}
\end{figure}

\begin{figure}[!h]
    \centering
    \includegraphics[width=8.6cm,height=4.6cm]{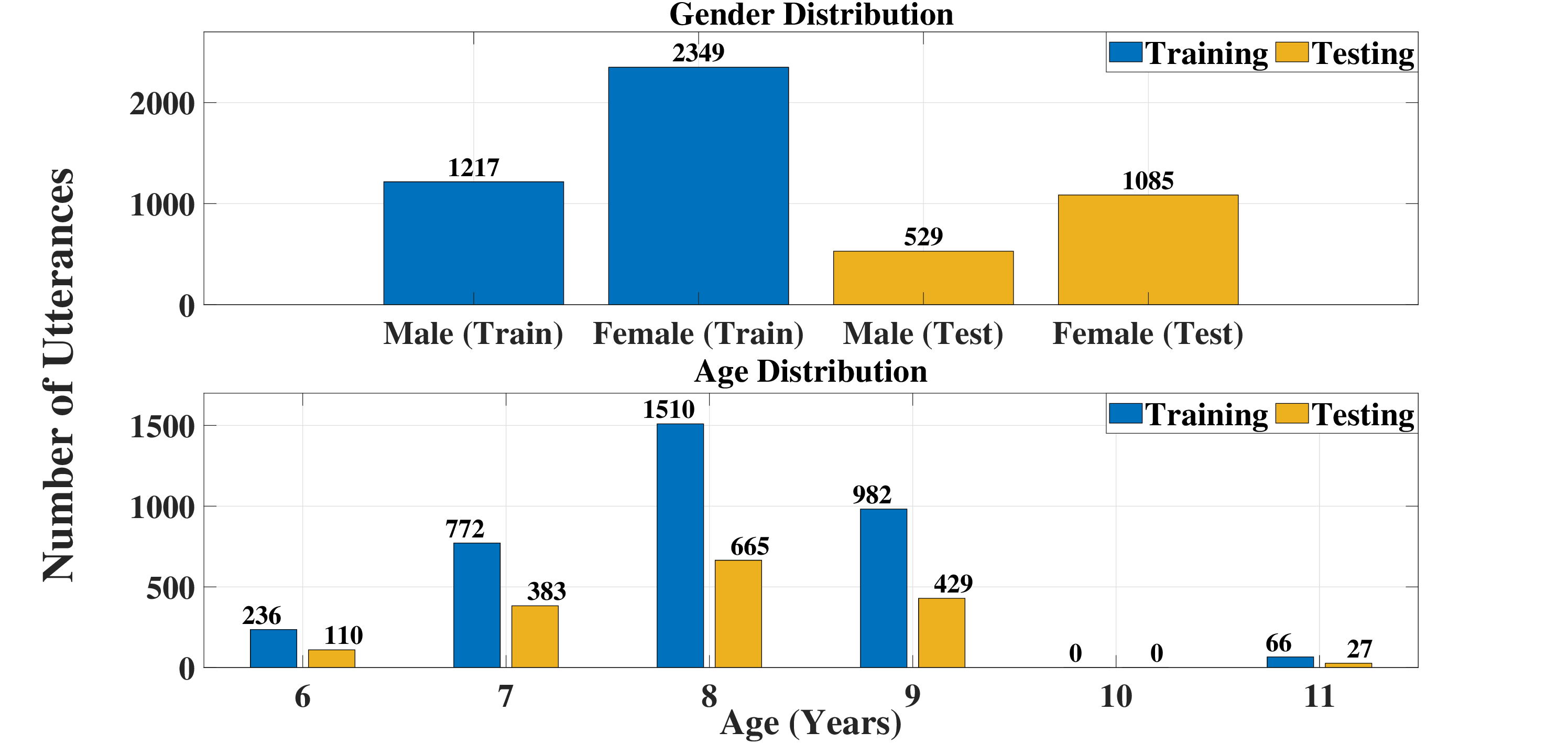} 
        
    \caption{Age and gender distribution of the CMU Kids dataset, showing training and testing splits by speaker gender and age. }
    \label{fig:data_dist_cmu}
\end{figure}

PFSTAR comprises read speech recorded in quiet environments, whereas CMU Kids includes both read and spontaneous utterances with moderate background noise. Notably, PFSTAR utterances are longer (avg. 41.32 sec vs. 6.28 sec in CMU Kids), which explains its higher total duration despite fewer utterances. Figures \ref{fig:data_dist_PFSTAR} and \ref{fig:data_dist_cmu} illustrates the dataset distributions used for age-wise, gender-wise, and cross-dataset fine-tuning. PFSTAR is divided into two age groups (4-8 and 9-14 years) and categorized by gender, while CMU Kids is split into two age groups (6-8 and 9-11 years) with balanced gender representation. The broader age range in PFSTAR introduces greater variability in speech patterns, whereas the balanced gender distribution in CMU Kids helps mitigate bias in gender-specific fine-tuning.

\vspace{-5pt}

\vspace{-10pt}

\subsection{Experimental Setup}
\label{ssec:ExperimentalSetup}

This section describes the SSL models used and the fine-tuning.

\subsubsection{Self Supervised Learning (SSL) Models}

We employed three state-of-the-art SSL models: Wav2Vec2-Large-960h-lv60-self, HuBERT-Large-LS960-ft, and WavLM-Large, hereafter referred to as Wav2Vec2, HuBERT, and WavLM, respectively. All three models are pre-trained on large-scale unlabeled audio data to learn robust speech representations, making them highly effective for ASR tasks. They share similar architectures with 25 hidden layers and a feature size of 1024: the initial CNN layer extracts features from raw audio, and the subsequent 24 Transformer layers capture contextual information.

Specifically, Wav2Vec2 was pre-trained on 60,000 hours of unlabeled data and fine-tuned on 960 hours of labeled data, while HuBERT used the same unlabeled corpus with a masked prediction strategy. WavLM was pre-trained on 94,000 hours from diverse sources and fine-tuned on 960 hours of labeled data.

The architecture processes raw audio through a CNN-based feature encoder, then passes the resulting feature vectors to a Transformer-based context network. A key component is the masking mechanism within the Transformer, where segments of the speech features are randomly masked and predicted from surrounding context, facilitating effective self-supervised learning. Each model employs a different loss function: Wav2Vec2 uses contrastive loss, HuBERT utilizes masked prediction loss with cluster assignments, and WavLM adopts noise-robust loss to handle speech variations. Despite these differences, the overall architecture remains consistent, enabling effective and generalizable speech representations.

\subsubsection{Fine-Tuning SSL Models for ASR}
To fine-tune the SSL models on children’s speech data, we followed the framework depicted in Figure \ref{fig:wav2vec2}. Fine-tuning was performed on both the overall dataset and on specific subsets defined by age (i.e., 4-8 and 9-14 years for PFSTAR; 6-8 and 9-11 years for CMU Kids) and gender. Each fine-tuned model was evaluated on multiple test sets, including the complete test set and its corresponding age- and gender-specific subsets.

For consistency across experiments, we used a fixed learning rate of 1e-4 and a weight decay of 0.005 to prevent overfitting. The models were trained with gradient checkpointing to optimize memory usage and improve generalization. A comprehensive vocabulary covering all transcription characters was built, and Connectionist Temporal Classification (CTC) loss was applied to align predicted sequences with the input speech. Evaluations employed greedy search decoding without an external language model to assess the intrinsic performance of the models.

\section{Results and Discussion}

This section presents a comprehensive analysis of the experimental results across various settings. Sections~\ref{sec:Baseline}-\ref{sec:crossdb} cover the baseline (zero-shot) performance, age group-specific and gender-specific fine-tuning outcomes, full-dataset fine-tuning improvements, and cross-dataset evaluations, examining performance trends, underlying factors, and implications for robust children’s ASR.

\subsection{Baseline Zero-Shot Results}
\label{sec:Baseline}
The baseline performance of three SSL models, Wav2Vec2, HuBERT, and WavLM was evaluated in a zero-shot setting on the PFSTAR and CMU Kids datasets. As shown in Table~\ref{tab:baseline}, Wav2Vec2 achieved the lowest WERs (10.65\% on PFSTAR and 22.37\% on CMU Kids), with HuBERT yielding similar results on PFSTAR (10.67\%) but a higher WER (24.24\%) on CMU Kids. In contrast, WavLM produced significantly higher error rates (25.42\% on PFSTAR and 34.25\%), suggesting that its pretraining objectives may not optimally capture the acoustic characteristics of children’s speech. WER is consistently higher for CMU Kids, likely due to its shorter utterances, suggesting that these SSL models struggle with shorter utterance lengths.

\begin{table}[!htbp]
    \centering
    \caption{WER(\%) from zero-shot decoding using three SSL models on the PFSTAR and CMU Kids datasets.}
    \renewcommand{\arraystretch}{1.2}
    \setlength{\tabcolsep}{8pt}
    \resizebox{8cm}{!}{
    \begin{tabular}{lccc}
    \hline
    \textbf{Model} & \textbf{Librispeech (Clean)} & \textbf{PFSTAR} & \textbf{CMU Kids} \\
    \hline
    Wav2Vec2 & 1.90 &\textbf{10.65} & \textbf{22.37} \\
    HuBERT   & 1.90 & 10.67 & 24.24 \\
    WavLM    & - & 25.42 & 34.25 \\
    \hline
    \end{tabular}}
    \label{tab:baseline}
\end{table}

\begin{table}[!t]
    \centering
    \caption{Zero-shot WER(\%) broken down by age group and gender-wise on the PFSTAR and CMU Kids datasets.}
    \renewcommand{\arraystretch}{1.5}
    \resizebox{8.6cm}{!}{
    \begin{tabular}{lcccccccc}
    \hline
    \multirow{3}{*}{\textbf{Model}} & \multicolumn{4}{c}{\textbf{PFSTAR}} & \multicolumn{4}{c}{\textbf{CMU Kids}} \\
    \cline{2-9}
    & \multicolumn{2}{c}{\textbf{Age Group}} & \multicolumn{2}{c}{\textbf{Gender}} 
    & \multicolumn{2}{c}{\textbf{Age Group}} & \multicolumn{2}{c}{\textbf{Gender}} \\
    \cline{2-9}
    & \textbf{4-8} & \textbf{9-14} & \textbf{Male} & \textbf{Female} 
    & \textbf{6-8} & \textbf{9-11} & \textbf{Male} & \textbf{Female} \\
    \hline
    Wav2Vec2 & \textbf{12.43} & 7.36 & \textbf{8.06} & 11.45  & \textbf{24.58} & \textbf{17.77} & \textbf{22.40} & \textbf{22.36}  \\
    HuBERT   & 13.61 & \textbf{6.91} & 8.16 & \textbf{11.40} & 27.03 & 18.57 & 25.62 & 23.51  \\
    WavLM    & 31.24 & 19.62 & 23.77 & 25.59 & 37.76 & 26.96 & 34.33 & 34.20  \\
    \hline
    \end{tabular}}
    \label{tab:baseline_age_gender}
\end{table}

An age-wise breakdown in Table~\ref{tab:baseline_age_gender} reveals that younger age groups consistently incur higher WERs than older groups. For PFSTAR, the 4-8 years group shows WERs approximately 5\% (Wav2Vec2) to 6.7\% (WavLM) higher than the 9-14 years group. Similar trends are observed in CMU Kids, where the 6-8 years group under performs the 9-11 years group by 6.81\% (HuBERT) and 10.8\% (WavLM). Gender-wise, PFSTAR shows male subsets outperforming female ones (e.g., an average difference of 3.39\% for Wav2Vec2), while CMU Kids exhibits minimal differences. Table~\ref{tab:baseline} further compares these results to Librispeech (adult speech), where all models achieve near state-of-the-art performance (e.g., 2.1\% for Wav2Vec2), confirming that the degradation on children’s speech arises from acoustic mismatches.

\subsection{Age Group-wise Fine-Tuning Results}
\label{sec:age_wise}

Table~\ref{tab:fine-tuning_results_age} shows that fine-tuning on younger children’s data yields models that generalize better to older age groups. In PFSTAR, models fine-tuned on the 4-8 years group achieve lower WERs on the 9-14 years test set (e.g., HuBERT: 7.13\% vs. 8.67\%) compared to fine-tuning on the older group. Similarly, in CMU Kids, Wav2Vec2 fine-tuned on the 6-8 years group achieves a WER of 7.47\% on the 9-11 years set, whereas fine-tuning on the older group results in a WER of 11.99\%. These trends suggest that greater acoustic variability in younger speech enables models to learn more robust representations. Conversely, models fine-tuned on older children generalize less effectively to younger children.

\begin{table}[!h]
    \centering
    \caption{WER(\%) achieved by fine-tuning SSL models across different age groups within the PFSTAR and CMU Kids datasets.}
    \renewcommand{\arraystretch}{1.2}
    \setlength{\tabcolsep}{8pt}
    \footnotesize
    \resizebox{8.6cm}{!}{
    \begin{tabular}{lcccccc}
        \hline
        \multicolumn{7}{c}{\textbf{PFSTAR Dataset}} \\
        \hline
        \multirow{2}{*}{\textbf{Model}} & \multicolumn{3}{c}{\textbf{Training Age Group: 4-8}} & \multicolumn{3}{c}{\textbf{Training Age Group: 9-14}} \\
        \cline{2-7}
         & \textbf{4-8 (Test)} & \textbf{9-14 (Test)} & \textbf{Average} & \textbf{4-8 (Test)} & \textbf{9-14 (Test)} & \textbf{Average} \\
        \hline
        Wav2Vec2 & 8.15 & 8.19 & 8.17 & \textbf{8.62} & \textbf{6.62} & \textbf{7.62} \\
        HuBERT   & \textbf{7.97} & \textbf{7.13} & \textbf{7.55} & 8.67 & 6.93 & 7.80 \\
        WavLM    & 8.34 & 9.45 & 8.89 & 7.87 & 7.57 & 7.72 \\
        \hline
        \multicolumn{7}{c}{\textbf{CMU Kids Dataset}} \\
        \hline
        \multirow{2}{*}{\textbf{Model}} & \multicolumn{3}{c}{\textbf{Training Age Group: 6-8}} & \multicolumn{3}{c}{\textbf{Training Age Group: 9-11}} \\
        \cline{2-7}
         & \textbf{6-8 (Test)} & \textbf{9-11 (Test)} & \textbf{Average} & \textbf{6-8 (Test)} & \textbf{9-11 (Test)} & \textbf{Average} \\
        \hline
        Wav2Vec2 & 3.94 & \textbf{7.47} & \textbf{5.70} & \textbf{11.99} & \textbf{4.06} & \textbf{8.02} \\
        HuBERT   & \textbf{3.35} & 8.55 & 5.95 & 12.00 & 4.08 & 8.04 \\
        WavLM    & 2.40 & 10.72 & 6.56 & 12.33 & 4.45 & 8.39 \\
        \hline
    \end{tabular}}
    \label{tab:fine-tuning_results_age}
\end{table}

\begin{table}[!h]
    \centering
    \caption{WER(\%) achieved by fine-tuning SSL models across gender groups within the PFSTAR and CMU Kids datasets.}
    \renewcommand{\arraystretch}{1.2}
    \footnotesize
    \resizebox{8.6cm}{!}{
    \begin{tabular}{lcccccc}
        \hline
        \multicolumn{7}{c}{\textbf{PFSTAR Dataset}} \\
        \hline
        \multirow{2}{*}{\textbf{Model}} 
        & \multicolumn{3}{c}{\textbf{Training Gender: Male}} 
        & \multicolumn{3}{c}{\textbf{Training Gender: Female}} \\
        \cline{2-7}
         & \textbf{Male (Test)} & \textbf{Female (Test)} & \textbf{Average} & \textbf{Male (Test)} & \textbf{Female (Test)} & \textbf{Average} \\
        \hline
        Wav2Vec2 & \textbf{6.09} & \textbf{10.65} & \textbf{8.37} & \textbf{5.47} & \textbf{8.50} & \textbf{6.99} \\
        HuBERT   & 6.94 & 10.70 & 8.82 & 6.81 & 10.18 & 8.50 \\
        WavLM    & 7.73 & 10.98 & 9.36 & 7.56 & 9.57 & 8.57 \\
        \hline
        \multicolumn{7}{c}{\textbf{CMU Kids Dataset}} \\
        \hline
        \multirow{2}{*}{\textbf{Model}} 
        & \multicolumn{3}{c}{\textbf{Training Gender: Male}} 
        & \multicolumn{3}{c}{\textbf{Training Gender: Female}} \\
        \cline{2-7}
         & \textbf{Male (Test)} & \textbf{Female (Test)} & \textbf{Average}
         & \textbf{Male (Test)} & \textbf{Female (Test)} & \textbf{Average} \\
        \hline
        Wav2Vec2 & \textbf{6.82} & \textbf{7.06} & \textbf{6.94} & 9.34 & 4.51 & 6.93 \\
        HuBERT   & 7.70 & 13.72 & 10.71 & \textbf{7.89} & \textbf{2.92} & \textbf{5.41} \\
        WavLM    & 5.54 & 9.59 & 7.57 & 8.28 & 3.07 & 5.68 \\
        \hline
    \end{tabular}}
    \label{tab:fine-tuning_results_gender}
\end{table}

\subsection{Gender-wise Fine-Tuning Results}
\label{sec:gender_wise}

Table~\ref{tab:fine-tuning_results_gender} summarizes the outcomes of gender-specific fine-tuning. In PFSTAR, models fine-tuned on male data generally achieve lower WERs on female test sets (e.g., Wav2Vec2: 8.37\% vs. 6.99\% average), with an average difference of 2-3\%. In CMU Kids, the gap is smaller (approximately 1-2\%), likely due to its balanced gender distribution. Notably, WavLM shows the largest relative improvement when generalized from male to female, while Wav2Vec2 maintains stable performance across genders. These results suggest that models fine-tuned on male speech perform better across genders, revealing a pretraining bias favoring male speech. However, this bias is reduced when fine-tuning on gender-balanced datasets, such as CMU Kids.

\subsection{Fine-Tuning on Entire Dataset}
\label{sec:fine-tuning}

Fine-tuning on the entire training dataset yields significant improvements in WER across both datasets, as shown in Table~\ref{tab:table5}. Wav2vec2 yeilds the best results on the overall test sets with WERs of 7.70\% and 5.43\% for the PFSTAR and CMU Kids dataset respectively. However WavLM shows the largest relative improvement when fine-tuned on the overall datasets. 

\begin{table}[!htbp]
    \centering
    \caption{WER(\%) achieved by fine-tuning three SSL models on the PFSTAR and CMU Kids datasets, including baseline and relative improvement (Rel. Imp.) over the baseline.}
    \renewcommand{\arraystretch}{1.2}
    \footnotesize
    \resizebox{8.6cm}{!}{
    \begin{tabular}{lcccccc}
        \hline
        \multirow{2}{*}{\textbf{Model}} & \multicolumn{3}{c}{\textbf{PFSTAR}} 
        & \multicolumn{3}{c}{\textbf{CMU Kids}} \\
        \cline{2-7}
         & \textbf{Baseline} & \textbf{Fine-Tuned} & \textbf{Rel. Imp.} 
         & \textbf{Baseline} & \textbf{Fine-Tuned} & \textbf{Rel. Imp.} \\
        \hline
        Wav2Vec2 & \textbf{10.65} & \textbf{7.70} & 27.7 & \textbf{22.37} & \textbf{5.43} & 75.7 \\
        HuBERT   & 10.67 & 7.84 & 26.5 & 24.24 & 5.96 & 75.4 \\
        WavLM    & 25.42 & 8.08 & 68.2 & 34.25 & 4.99 & 85.4 \\
        \hline
    \end{tabular}}
    \label{tab:table5}
\end{table}

When evaluated on the corresponding age and gender-specific test sets, for PFSTAR, the fine-tuned Wav2Vec2 achieves a WER of 6.94\% for the 4-8 years age group and 6.24\% for the 9-14 years group, while HuBERT and WavLM show comparable gains. In the CMU Kids dataset, Wav2Vec2 achieves 5.85\% and 4.52\% for the 6-8 and 9-11 years age groups, respectively. Table~\ref{tab:age_gender} further illustrates that while fine-tuning on the full dataset significantly reduces WER across all groups, age- and gender-specific fine-tuning provides additional benefits for specific subgroups, highlighting the importance of targeted adaptation.

\begin{table}[!htbp]
    \centering
    \caption{WER(\%) achieved by fine-tuning SSL models on the PFSTAR and CMU Kids datasets, evaluated on corresponding age and gender test sets.}
    \renewcommand{\arraystretch}{1.2}
    \resizebox{7.8cm}{!}{
    \begin{tabular}{lcccccccc}
    \hline
    \multirow{3}{*}{\textbf{Model}} & \multicolumn{4}{c}{\textbf{PFSTAR}} & \multicolumn{4}{c}{\textbf{CMU Kids}} \\
    \cline{2-9}
    & \multicolumn{2}{c}{\textbf{Age Group}} & \multicolumn{2}{c}{\textbf{Gender}} 
    & \multicolumn{2}{c}{\textbf{Age Group}} & \multicolumn{2}{c}{\textbf{Gender}} \\
    \cline{2-9}
    & \textbf{4-8} & \textbf{9-14} & \textbf{Male} & \textbf{Female} 
    & \textbf{6-8} & \textbf{9-11} & \textbf{Male} & \textbf{Female} \\
    \hline
    Wav2Vec2 & 6.94 & \textbf{6.24} & \textbf{5.16} & 8.50 & 5.85 & \textbf{4.52} & 5.68 & 5.30 \\
    HuBERT   & 6.94 & 6.55 & 5.57 & \textbf{8.27} & 6.34 & 5.16 & 6.00 & 5.94 \\
    WavLM    & \textbf{6.51} & 7.23 & 5.75 & 8.31 & \textbf{5.08} & 4.79 & \textbf{5.38} & \textbf{4.77} \\
    \hline
    \end{tabular}}
    \label{tab:age_gender}
\end{table}

\subsection{Cross Dataset Evaluation}
\label{sec:crossdb}

Cross-dataset fine-tuning further highlights the challenge of generalizing across datasets with distinct accents, demographics, and recording conditions. Tables~\ref{tab:crossdb_cmu_kids} and~\ref{tab:crossdb_PFSTAR} summarize the performance when models fine-tuned on one dataset are evaluated on the other. When fine-tuned on PFSTAR (British English) and tested on CMU Kids (American English), WavLM and HuBERT show WER increases of approximately 8-10\% due to accent, vocabulary mismatches and greater acoustic variability, whereas Wav2Vec2 exhibits relatively smaller increases in WER. Similarly, models fine-tuned on CMU Kids experience notable degradation when tested on PFSTAR.

These results shows that models fine-tuned on one dataset struggle to generalize to another, with WER increasing significantly compared to the baseline due to accent and demographic mismatches. 

\begin{table}[!htbp]
    \centering
    \caption{WER(\%) achieved by fine-tuning SSL models on PFSTAR and testing on CMU Kids, for overall and age/gender-specific subsets.}
    \renewcommand{\arraystretch}{1.2}
    \resizebox{8.6cm}{!}{
    \begin{tabular}{lccccc}
    \hline
        \multirow{2}{*}{\textbf{Model}} & \multicolumn{5}{c}{\textbf{Testing Sets CMU Kids}} \\
        \cline{2-6}
        & \textbf{Overall} & \textbf{Age 6-8} & \textbf{Age 9-11} & \textbf{Male} & \textbf{Female} \\
        \hline
        Wav2Vec2 & 34.37 & 37.21 & 28.34 & 33.94 & 34.59 \\
        HuBERT   & 37.37 & 39.83 & 32.15 & 36.84 & 37.65 \\
        WavLM    & 46.61 & 49.30 & 42.33 & 46.13 & 46.81 \\
        \hline
    \end{tabular}}
    \label{tab:crossdb_cmu_kids}
\end{table}

\begin{table}[!htbp]
    \centering
    \caption{WER(\%) achieved by fine-tuning SSL models on CMU Kids and testing on PFSTAR, for overall and age/gender-specific subsets.}
    \renewcommand{\arraystretch}{1.2}
    \resizebox{8.6cm}{!}{
    \begin{tabular}{cccccc}
    \hline
        \multirow{2}{*}{\textbf{Model}} & \multicolumn{5}{c}{\textbf{Testing Sets PFSTAR}} \\
        \cline{2-6}
        & \textbf{Overall} & \textbf{Age 4-8} & \textbf{Age 9-14} & \textbf{Male} & \textbf{Female} \\
        \hline
        Wav2Vec2 & 63.71 & 64.95 & 62.81 & 68.02 & 70.29 \\
        HuBERT   & 79.51 & 80.83 & 78.54 & 79.17 & 79.96 \\
        WavLM    & 75.32 & 79.66 & 72.16 & 74.45 & 76.51 \\
        \hline
    \end{tabular}}
    \label{tab:crossdb_PFSTAR}
\end{table}




Overall, our key findings are:
\begin{itemize}
    \item \textbf{Fine-tuning closes the gap:} Although zero-shot ASR on children’s speech suffers from acoustic mismatches, targeted fine-tuning yields substantial WER reductions.
    \item \textbf{Subgroup adaptation matters:} Age-specific fine-tuning delivers the greatest gains for younger speakers, and gender-specific training helps counteract pretraining biases.
    \item \textbf{Cross-corpus brittleness:} Transferring between datasets incurs large performance drops due to accent, vocabulary, and demographic mismatches highlighting the need for more diverse, representative pretraining data.
\end{itemize}

\section{Conclusion}

We evaluated SSL models for children’s ASR using three fine-tuning strategies (age-specific, gender-specific, and cross-dataset) on the PFSTAR and CMU Kids corpora. Our findings highlight challenges in cross-domain generalization, pretraining biases, and dataset mismatches. Fine-tuning on younger children’s speech improved generalization to older children-likely because younger speech has greater acoustic variability-whereas models trained on older children generalized poorly to younger speakers. Gender-specific fine-tuning revealed a bias favoring male speech: models trained on male voices performed relatively better, but balanced gender training sets mitigated this bias, underscoring the importance of diverse gender representation. Cross-dataset evaluation showed significant WER degradation due to accent, vocabulary, and recording differences, emphasizing the need for more diverse pretraining data. We also observed that baseline WER was lower for the longer PFSTAR utterances than for the shorter CMU Kids utterances, suggesting that these pre trained ASR models struggle more with short, acoustically variable speech. Overall, these results indicate that child-centric ASR systems benefit from targeted fine-tuning on specific subgroups and from pretraining on diverse, child-relevant data.

\bibliographystyle{IEEEtran}
\bibliography{mybib,refs}

\end{document}